\documentclass[aps,prl,unsortedaddress,groupedaddress,preprintnumbers,amsmath,amssymb,twocolumn,floatfix]{revtex4}

\usepackage{graphicx}% Include figure files
\usepackage{subfigure}
\usepackage{epsfig}
\usepackage[paperwidth=210mm,paperheight=297mm,centering,hmargin=2cm,vmargin=2.5cm]{geometry}% http://ctan.org/pkg/geometry
\usepackage{placeins}
\usepackage{float}
\usepackage{graphicx}% http://ctan.org/pkg/graphicx
\usepackage{dcolumn}% Align table columns on decimal point
\usepackage{bm}% bold math
\usepackage{hyperref}% add hypertext capabilities
\everymath{\displaystyle}
\usepackage{hyphenat}
\emergencystretch=\maxdimen
\hyphenchar\font=-1
\begin{document}

\title{Anomalous Non-linear Optical Response Of Graphene Near Phonon Resonances}

%\title{Interference Effects on the Third Order Optical Nonlinearity of Graphene and h-BN}

\author{Lucas Lafet\'a$^{1}$}
\author{Alisson Cardore$^{1}$}
\author{Thiago G. Mendes S\'a$^{1}$}
\author{Kenji Watanabe$^{2}$}
\author{Takashi Taniguchi$^{2}$}
\author{Leonardo C. Campos$^{1}$}
\author{Ado Jorio$^{1}$}
\author{Leandro M. Malard$^{1}$}
\email{lmalard@fisica.ufmg.br}
\address{$^{1}$Departamento de F\'isica, Universidade Federal de Minas Gerais, Belo Horizonte, MG 31270-901, Brazil}
\address{$^{2}$National Institute for Materials Science (NIMS) | 1-2-1 Sengen, Tsukuba-city Ibaraki 305-0047 Japan}

\date{\today}

\begin{abstract}

In this work we probe the third-order non-linear optical property of graphene, hexagonal boron nitride and their heterostructure by the use of coherent anti-Stokes Raman Spectroscopy. When the energy difference of the two input fields match the phonon energy, the anti-Stokes emission intensity is enhanced in h-BN, as usually expected while for graphene a anomalous decrease is observed. This behaviour can be understood in terms of q coupling between the electronic continuum and a discrete phonon state. We have also measured a graphene/h-BN heterostructure and demonstrate that the anomalous effect in graphene dominates the heterostructure optical response.
\end{abstract}

\pacs{Valid PACS appear here}

\maketitle

Two-dimensional materials like graphene, hexagonal boron nitride (h-BN) and heterostructures exibit novel physical properties and promisses different applications in electronics and photonics \cite{novoselov2011,bonaccorso2010,geimreview,xia2014}. Different non-linear optical phenomena like second-, third-harmonic generation and four wave mixing (FWM) \cite{franken1961,shen,boyd} can be quite strong in these materials \cite{hendry2010,lukas1,hong2014,yilei2013,kumar2013,malardshg,jena2015,matos2017}. However, the interpretation of the nonlinear optical response is strongly affected by electronic and phonon resonances \cite{porri14,zenobi,huangreview,potma1}, therefore the knowledge of the interplay between these resonances is desirable. Here we measured the third order optical emission by the degenerated four wave mixing emission of graphene, h-BN and their heterostructure near phonon resonances. We show that while the FWM signal in h-BN shows the expected enhancement, in graphene the signal is decreased exactly at the phonon resonance. These results are explained in terms of interference effects between the electronic continuum and discrete phonon states for these two different materials. We also show that this unusual effect in graphene dominates the optical response in the graphene/h-BN heterostructure. 

Four wave mixing is a third-order non-linear optical phenomena, where three frequencies are combined to generate a forth \cite{boyd}. In this work we are restricted to the case of degenerate FWM, where two photons of frequency $\omega_{1}$ combines with a photon of $\omega_{2}$ at the material and generate the emission of another photon with frequency $\omega_{4}$. The energy conservation in this case is given by $\hbar\omega_{4}=2\hbar\omega_{1}-\hbar\omega_{2}$. Hendry et al. \cite{hendry2010} have measured the FWM intensity in graphene as a function of the pump laser energy and its third order optical non linear optical property was characterized. Also different theoretical works have calculated the third-order optical conductivity of graphene \cite{Mikhailov2012924,sipe1,polini1,mikhailov16}, showing the importance of different physical quantities like Fermi energy or temperature. However these works did not treat the problem of the third order optical nonlinearity near phonon resonances. The so-called coherent anti-Stokes Raman spectroscopy (CARS) is a special case of FWM when the energy difference between $\hbar\omega_{1}$ and $\hbar\omega_{2}$ matches a phonon energy ($\hbar\omega_{ph}$), then $\omega_{4}$ corresponds exactly to the anti-Stokes frequency in Raman scattering. In general, when the energy condition $\hbar\omega_{1}-\hbar\omega_{2}=\hbar\omega_{ph}$ is satisfied, the $\omega_{4}$ amplitude is enhanced \cite{porri14,zenobi,huangreview,potma1}.  

%Here in this work we have studied the amplitude dependence of the $\omega_{4}$ emission in two-dimensional materials near phonon resonances. We have found that the CARS spectrum is strongly influenced by the different types of electronic band structure in 2D materiais like graphene and h-BN.%

In order to study the CARS phenomenon in graphene and h-BN, flakes were prepared by micro-mechanical cleavage of natural graphite or bulk h-BN in transparent quartz substrates (SPI Inc.). Monolayer graphene were located in the substrates by an optical microscope followed by Raman characterization, where the 2D Raman band in monolayer graphene was characterized by a single Lorentzian \cite{Malard200951}. The h-BN flakes used were fewlayers (10-20 layers). For the CARS experiment we have used an optical parametric oscillator system (APE PicoEmerald) with 6 ps pulse width and 76 MHz repetition rate. This laser system emits two collinear laser beams with frequencies $\omega_{1}$ and $\omega_{2}$. The frequency $\omega_{1}$ can be tuned between 720-960 nm in steps of 0.5 nm and the second $\omega_{2}$ is fixed at 1064 nm.. As Both laser beams are spatially and temporally overlapped and focused at the sample by a 60$\times$ and 0.95 N.A. objective. The backscattered signal is collected by the same objective, reflected by a beamsplitter (BS), filtered by a short pass (SP) at to remove the pump wavelengths and is directed to a single grating spectrometer equipped with a CCD camera [see schematics in Fig. \ref{fig1} (a)]. Raman spectra where acquired in the same setup, but using a 561 nm diode laser with a edge filter in front of the spectrometer [bottom panel in Figs. \ref{fig1} (b) and (c)].

\begingroup
\begin{figure}[!htp]
%\advance\leftskip-3cm
%\advance\rightskip-3cm
    \centering
    \includegraphics[scale=0.4]{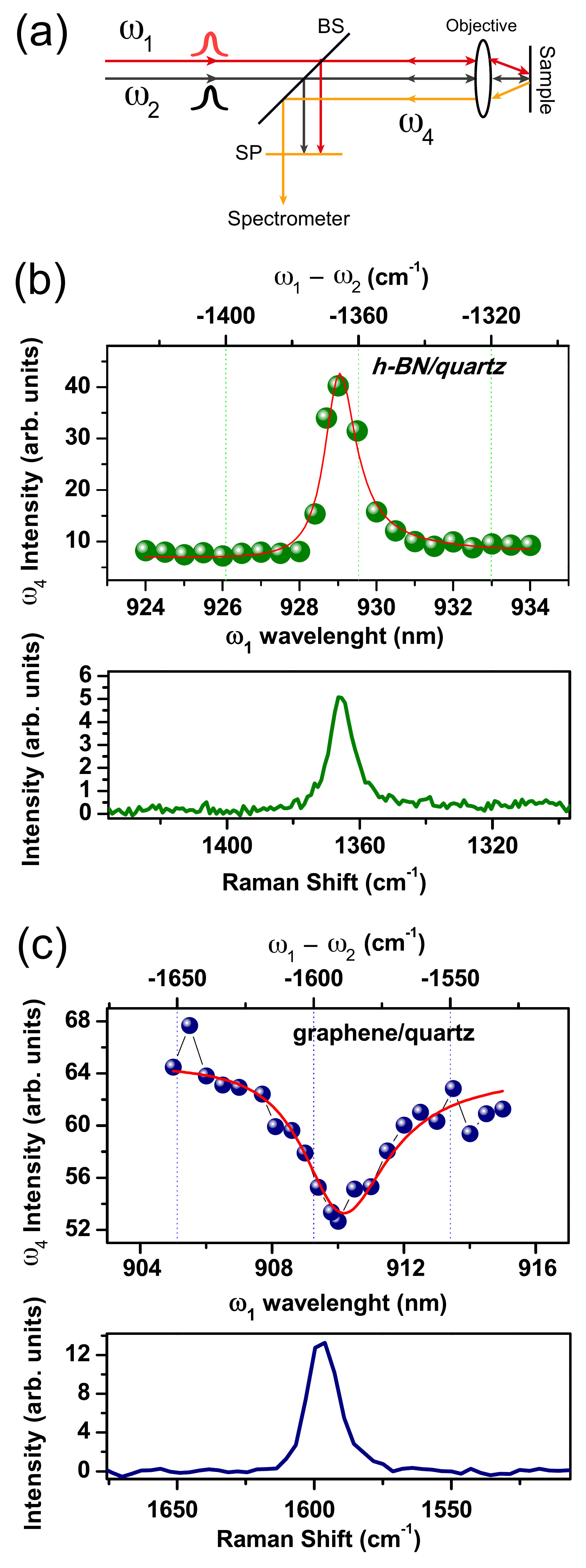} %size for onecolumn
    \caption{(Color Online)
    (a) Experimental setup used showing the two pump beams with frequency $\omega_{1}$ and $\omega_{2}$. (b) CARS intensity as a function of the $\omega_{1}$ pump wavelength (bottom scale) or $\hbar\omega_{2}-\hbar\omega_{1}$ in wavenumbers (top scale) for fewlayer h-BN. The solid red lines is the fit from the theory described in text. The graph bellow shows the Raman spectrum taken at the same energy range. (c) Same but for the monolayer graphene sample.}
    \label{fig1}
\end{figure}
\endgroup

Figure \ref{fig1} (b) shows the CARS spectrum ($\omega_{4}$ intensity as a function of $\omega_{1}$ wavelength) in fewlayer h-BN deposited on quartz substrate. The $\omega_{4}$ intensity is enhanced when the pump wavelength $\omega_{1}$ is around 929 nm. Converting the bottom scale to $\hbar\omega_{2}-\hbar\omega_{1}$ in wavenumbers (top scale), the enhancement in CARS intensity happens exactly at 1366 cm$^{-1}$, which corresponds to the doubly degenerated in plane optical phonon mode in h-BN. To further verify this assignment, the sample is measured by linear Raman spectroscopy as shown in the Raman spectrum at the bottom plot of Fig. \ref{fig1} (b).

Figure \ref{fig1} (c) shows the measurement of a monolayer graphene deposited on a similar quartz substrate, surprising the non-linear optical behaviour is the opposite, i.e. the CARS intensity decreases when the pump wavelength is $\sim$ 910 nm. Again, converting the bottom scale to $\hbar\omega_{2}-\hbar\omega_{1}$, we verify that the observed anti-resonance in the $\omega_{4}$ intensity is centered at the 1590 cm$^{-1}$ peak i.e. the energy of the doubly degenerated in plane optical phonon mode in graphene (G band). This assignment is confirmed by the Raman spectrum of the same monolayer graphene sample shown in Fig. \ref{fig1} (c), bottom plot.

The $\omega_{4}$ emission intensity (I$_{\omega_{4}}$) in a CARS process depends on the intensity of $\omega_{1}$ (I$_{\omega_{1}}$) and $\omega_{2}$ (I$_{\omega_{2}}$), and on the frequency-dependent third-order non-linear susceptibility $\chi^{(3)}(\omega)$ as $I_{\omega_{4}}\propto |\chi^{(3)}(\omega)|^{2} I_{\omega_{1}}^{2}I_{\omega_{2}}$ \cite{boyd,xiecars1}. Here the phase matching condition is ignored because the thickness of the 2D materials analysed here are much smaller than the pump wavelengths. In a simple analysis, where only electronic virtual states are present the $\chi^{(3)}(\omega)$ is a function with real and complex parts. The real function describes the so-called non-resonant background due to light absorption by virtual electronic states. The imaginary function describes a resonant state due to transition between virtual excited electronic states to a discrete phonon state \cite{boyd,xiecars1}. As shown before (\cite{xiecars1} and references therein) this simple analysis describes well the CARS spectrum for different transparent materials and molecules. However, such model can not explain the anti-resonance behaviour observed for graphene as shown in Fig. \ref{fig1} (c).

 %To understand the results shown in Fig. \ref{fig1}, one needs to calculate the dependence of the third order nonlinear with frequency explicitly, considering the different possible pathways of light absorption: between electronic and phonon states present in the material.%

\begin{figure}
\centering
\includegraphics[scale=1.7]{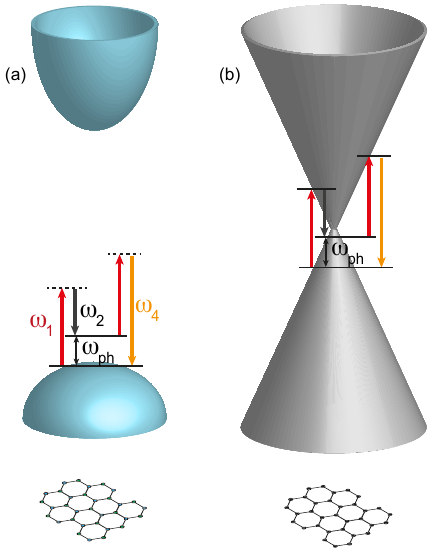}
\caption{(Color Online)
(a) h-BN band structure with a possible CARS process , where the two pump beams with frequencies $\omega_{1}$ and $\omega_{2}$ are combined to generate the emission of photons with frequency $\omega_{4}$ obyeing $\omega_{1}-\omega_{2}=\omega_{ph}$ . (b) The same for graphene.}
\label{fig2}
\end{figure}

The CARS process obeys $\omega_{4}=2\omega_{1}-\omega_{2}$, as shown in Fig. \ref{fig2}. h-BN is a insulator with a band gap energy higher than the laser energies used in this work \cite{zhen_hbn}, therefore the pump photon $\omega_{1}$ makes a first transition from the ground state to a virtual excited state, followed by a transition induced by the $\omega_{2}$ photon to a real phonon state [see Fig.\ref{fig2} (a)]. Another $\omega_{1}$ photon makes a transition from the phonon state to a virtual state and the $\omega_{4}$ photon is created by decaying from the second virtual state to the ground state. Therefore, for h-BN the CARS process connects only virtual electronic states and a real phonon state when the energy condition $\hbar\omega_{1}-\hbar\omega_{2}=E_{ph}$ is satisfied. This is the usual CARS process and it can be described by the $\chi^{(3)}(\omega)$ function as described in the previous paragraph.

The situation is different in monolayer graphene, which is a zero gap semiconductor, where the valence and conduction bands touch each other at the K and K$^{\prime}$ points of the Brillouin zone. In Fig. 2(b) the same CARS process is depicted, however in this case all the excited electronic states involved can be real, and resonance can be always achieved, not only in the four optical processes but also within the phonon $\omega_{ph}$ energy. Therefore, in the case of graphene, the presence of a continuum of electronic resonances cannot be ignored in the interpretation of the CARS spectrum.

CARS spectrum with pump lasers close to electronic transitions were studied in the past. Nestor {\it et al.} \cite{nestor1} observed that, by tuning the pump laser wavelength near the absorption peak of vitamin B$_{12}$, it was possible to observe a transition between a resonance to anti-resonance profile in the CARS spectra. Theoretical efforts have been make to treated this problem and multiple $\chi^{(3)}$ terms exists depending of resonant conditions \cite{lotem1,druet1,boyd}. For the case of graphene, different works have addressed the calculation of $\chi^{(3)}$ as function of energy, however the influence of discrete phonon states are still lacking in the literature\cite{Mikhailov2012924,sipe1,polini1,mikhailov16}. However, the $\chi^{(3)}$ energy dependence can be described by the phenomenological Fano lineshape \cite{fano1} in order to describe the CARS process. Such methodology has been applied to understand the FWM \cite{meier1995,shapiro2016} and CARS spectra of atomic systems at high energies \cite{tagkey1988ii}. The Fano lineshape can be written as a function of energy $E$ as:

\begin{equation}
I_{\omega_{4}}(E)=A\frac{[(E-E_{ph})+\gamma q]^{2}}{(E-E_{ph})^{2}+\gamma^{2}},
\label{eq1}
\end{equation}
where A is a proportionality constant, $E$ is the energy difference between the pump beams ($\hbar\omega_{1}-\hbar\omega_{2}$), $E_{ph}$ is the phonon energy, $\gamma$ the phonon state broadening and $q$ is a dimensionless parameter that gives the overall contribution between an electronic continuum or a discrete phonon state for the FWM intensity. If $|q|=1$ there is an equal weight contribution between the electronic continuum and discrete phonon state. If $|q|<<1$ there is a larger contribution to the electronic continuum and if $|q|>>1$ the contribution mainly comes from the discrete phonon state. When the phonon discrete state dominates ($|q|>>1$) there is a resonance lineshape and the profile is similar to the usual CARS profess. When the electronic continuum dominates ($|q|<<1$), however, there is a anti-resonance lineshape. We have used  Eq. \ref{eq1} to fit our experimental results in Fig. \ref{fig1}. For h-BN, the $q$ value found is $-6$ (discrete phonon state dominates) giving rise to a resonant behavior of the CARS spectrum at the phonon energy. On the other hand, the value found for $q$ in monolayer graphene is $0.09$ (continuum electronic states dominates) which leads to the anti-resonance CARS lineshape behaviour at the phonon energy. Based on this Fano analysis our results can now be understood: in graphene the electronic contributions are expected to play a major role due to the absence of an optical gap, i.e. there is a continuum of optical resonances. Therefore, in graphene the CARS intensity presents an anti-resonance behaviour exactly at the phonon energy. In contrast, in h-BN the optical gap is much larger then the energies used in our experiment, therefore we expect very low contribution from the electronic states, and the CARS intensity is a resonance peak located exactly at the phonon energy.

From the fitting we can also extract the phonon linewidth $\gamma$, which serves as an internal consistency analysis. For graphene $\gamma$ is found to be equal 12 cm$^{-1}$ is in agreement with value measured from Raman spectroscopy of 11 cm$^{-1}$ (Fig. \ref{fig1}). For h-BN we have found $\gamma=8$ cm$^{-1}$, which is close to our experimental resolution of 6 cm$^{-1}$, but in agreement with the value found for Raman spectroscopy of 9 cm$^{-1}$.

\begin{figure}
\centering
\includegraphics[scale=0.5]{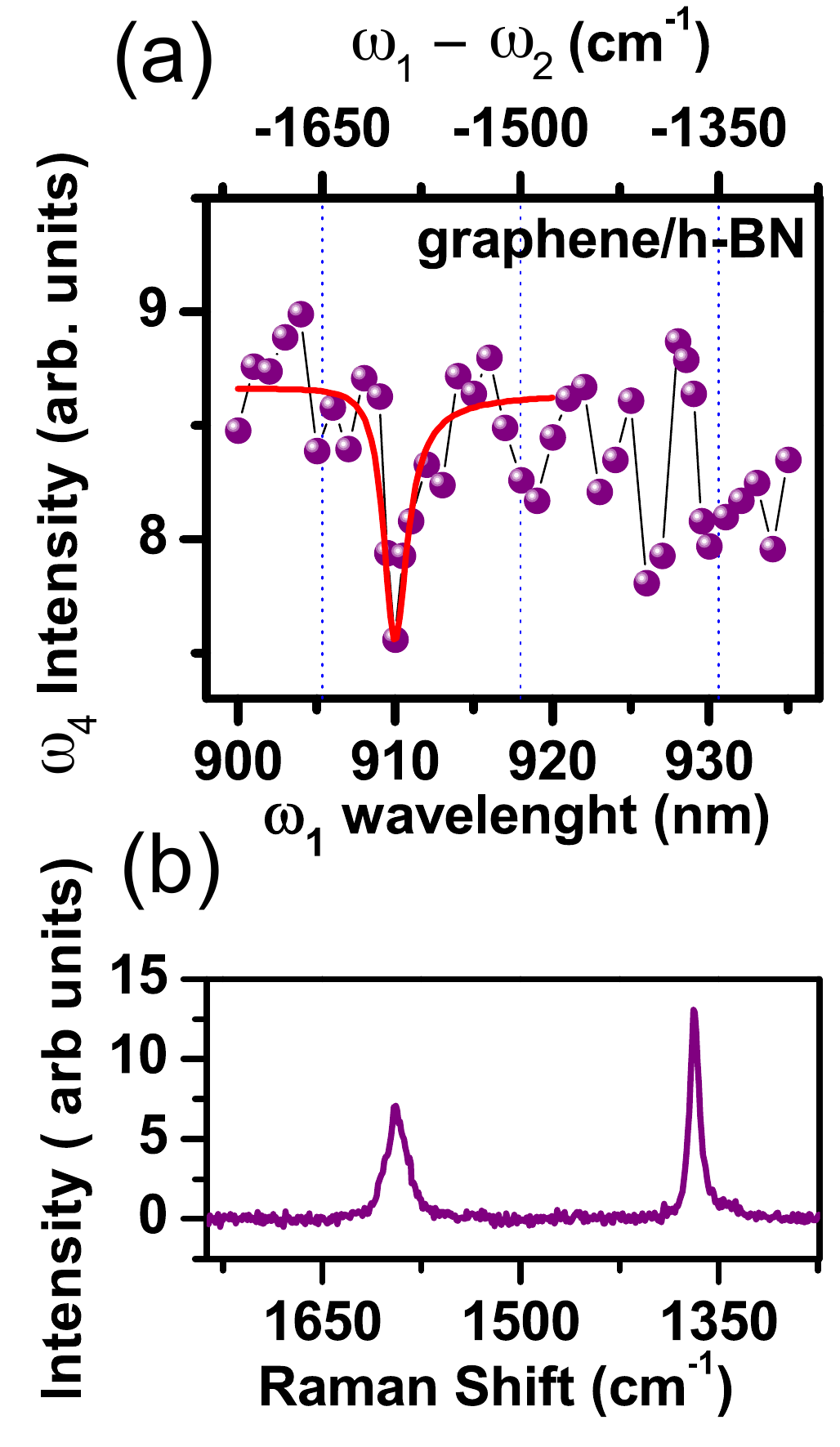}
\caption{(Color Online)
(a) CARS spectrum of a heterostructure formed by graphene on top of a fewlayer h-BN deposisted on Si/SiO$_{2}$. The solid red lines are the fitted data using Eq. \ref{eq1} (b) Raman spectrum of the same sample.}
\label{fig3}
\end{figure}

Finally we have built a graphene/h-BN heterostructure by transferring a monolayer graphene sample on top of fewlayer h-BN, following Ref. \cite{leocampos1}. In Fig. \ref{fig3}, the CARS spectrum for a graphene/h-BN heterostructure is shown. The anti-resonance lineshape can be seen at the graphene phonon energy, while the resonant lineshape at the h-BN phonon energy is not clearly resolved, despite the fact that the linear Raman spectrum (bottom plot) shows both phonon modes. Actually, the h-BN Raman peak (1366 cm$^{-1}$) in more intense than the graphene peak (1590 cm$^{-1}$). By using Eq. \ref{eq1} to fit the experimental CARS spectrum near the graphene phonon energy, we have found $q=0.1$, which is very similar to the case of graphene on top of quartz substrate, showing that the graphene signal completely dominates the nonlinear optical response of the heterostructure. Since graphene on top of boron nitride in know to be a very clean sample, with low electron doping and high mobility \cite{geimreview}, this result implies that the effect is robust in clean graphene samples and low doping levels. 

In conclusion, we have measured the third order optical non-linear property of graphene, h-BN and their heterostructure by coherent anti-Stokes spectroscopy. The CARS intensity as a function of energy was modeled by a Fano lineshape, which is shown to have good agreement with our experimental data. The observed anomalous anti-resonance behaviour for graphene and resonance behaviour for h-BN was explained in terms of strong contribution arising form the available continuum of electronic states in graphene. This third-order behaviour of graphene dominates the optical response of the heterostructure. We believe that our results provides new information for further development of theoretical works inteded to describe the third-order non-linear optical effects in these two-dimensional materials.

We acknowledge financial support from the Brazilian agencies CNPq, FAPEMIG, FINEP, INCT Medicina Molecular and Rede Brasileira de Instrumenta\c{c}\~ao CNPq.

\bibliography{references}

\end{document}